\documentclass{article} 
\usepackage{iclr2027_conference,times}

\usepackage{amsmath,amsfonts,bm}

\def\eqref#1{equation~\ref{#1}}

\def\1{\bm{1}}

\DeclareMathAlphabet{\mathsfit}{\encodingdefault}{\sfdefault}{m}{sl}
\SetMathAlphabet{\mathsfit}{bold}{\encodingdefault}{\sfdefault}{bx}{n}

\usepackage{hyperref}
\usepackage{url}
\usepackage{graphicx}

\title{MusicSTAR: Structure-Aware Temporal Attention Routing for Song Generation from a Single Long Caption}

\author{
\textbf{Fan Wang}$^{1,\dagger}$,
\textbf{Yutang Feng}$^{2,\dagger}$,
\textbf{Jiatao Chen}$^{2}$,
\textbf{Shun Lei}$^{1}$,
\textbf{Jiehui Luo}$^{1}$,
\textbf{Mingyu Zhao}$^{1}$,\\
\textbf{Dapeng Wu}$^{1}$,
\textbf{Zijian Lin}$^{1}$,
\textbf{Jinchao Zhang}$^{2,*}$,
\textbf{Zhiyong Wu}$^{1,*}$ \\[0.5em]
$^{1}$Shenzhen International Graduate School, Tsinghua University, Shenzhen, China \\
$^{2}$WeChat AI, Tencent, China \\[0.3em]
$^{\dagger}$Equal contribution. \quad
$^{*}$Corresponding authors. \\[0.3em]
\texttt{wang-f25@mails.tsinghua.edu.cn},
\texttt{yutangfeng@tencent.com}, \\
\texttt{dayerzhang@tencent.com},
\texttt{zywu@sz.tsinghua.edu.cn}
}

\iclrfinalcopy 
\begin{document}

\maketitle

\begin{abstract}
Controllable song generation requires specifying not only a song's overall characteristics, but also how they evolve over time. A single long caption offers a natural way to express this evolution by organizing global characteristics and ordered section-level descriptions into one structured temporal specification. Yet such a description does not by itself ensure reliable temporal control: during autoregressive generation, the model can access conditions from different sections, leading to temporal condition misrouting. Motivated by this observation, we propose MusicSTAR, a structure-aware framework for temporal attention routing from a single long caption. MusicSTAR factorizes section-level conditions and dynamically routes attention to the relevant section while preserving access to acoustic history, with a temporal scheduler adaptively controlling transitions between successive conditions without explicit timestamps or durations. We further introduce local-to-global preference optimization, progressing from section-level controllability to whole-song musicality. Experiments show that MusicSTAR substantially improves fine-grained temporal instruction following while maintaining strong overall song quality, demonstrating that effective temporal control depends not only on what is specified, but also on how that information is accessed during generation.
\end{abstract}

\section{Introduction}
Recent advances in song generation have substantially improved lyrics-conditioned and text-guided synthesis. Yet generating a compelling song is inherently a temporal problem: musical attributes such as energy, mood, vocals, instrumentation, rhythm, and sound style evolve across sections rather than remaining fixed throughout a song. Meaningful control therefore requires specifying not only a song's overall characteristics, but also how they change as the song unfolds.


Recent work has begun to address this challenge by making temporal structure explicit. SegTune~\cite{wang2026segtune} pairs a global prompt with segment-level descriptions and explicitly aligns each local condition with its temporal window during generation, using an LLM-based duration predictor to obtain timestamped representations. Muse~\cite{jiang2026muse} instead interleaves global and segment-level conditions within a conversational, segment-structured generation process, associating each segment with its lyrics and corresponding audio-token span. In both cases, temporal structure is imposed \emph{externally}, either by predicted alignment or by the generation format itself, which motivates a complementary question: \emph{can the temporal evolution of an entire song be expressed in a single long caption and resolved internally, within one unified autoregressive generation process?}


\begin{figure}[t]
\begin{center}
\includegraphics[width=\linewidth]{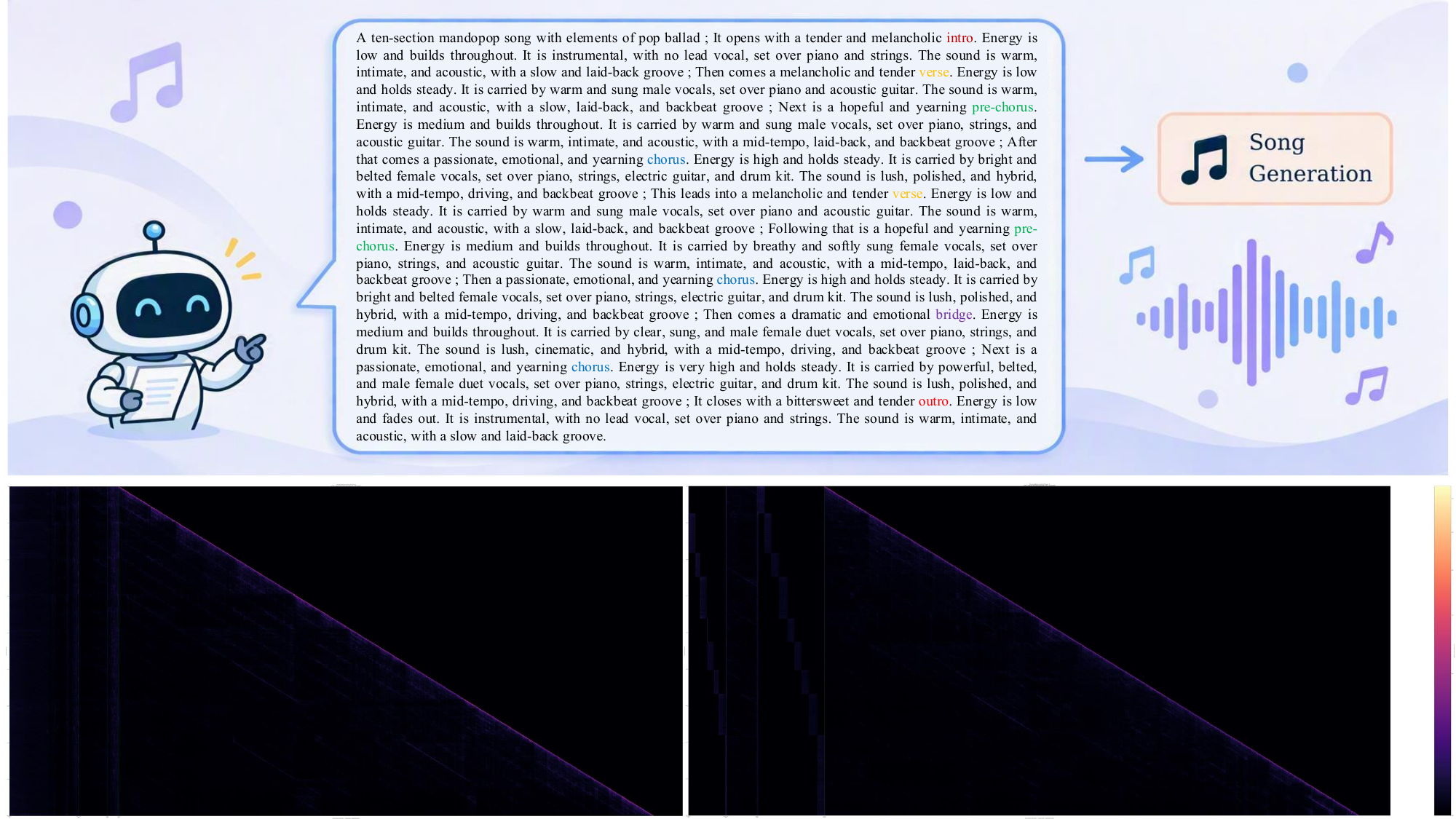}
\end{center}
\caption{Motivation for MusicSTAR. Top: a structured single long caption of a ten-section song. Bottom: attention over input tokens (x-axis) as generation proceeds (y-axis), for the baseline (left) and MusicSTAR (right). The baseline's caption attention remains diffuse, while MusicSTAR's advances diagonally along the section order.}
\label{fig:motivation}
\end{figure}

We explore this formulation with a structured single long caption that combines a global description with an ordered sequence of section-level descriptions, each rendered in natural language from a normalized set of user-facing temporal attributes. The caption thus describes salient musical characteristics as an ordered evolution rather than a static summary, providing a temporally complete control specification in a single textual condition without explicit timestamps or durations.


The key question is then how an autoregressive model actually uses such a caption: at every generation step the full caption remains accessible, although only the condition relevant to the current musical state should dominate. To examine this, we visualize the inference-time attention of LeVo~\cite{lei2026levo}, a strong open-source autoregressive song generation model, and compare it with MusicSTAR as shown in Fig.~\ref{fig:motivation}. The baseline's attention over the caption remains broadly distributed across generation stages, whereas MusicSTAR exhibits a clear diagonal pattern that advances through the ordered section descriptions as the song unfolds. We refer to this mismatch as \emph{temporal condition misrouting}: the specification is complete, but its access is not temporally selective. Fine-grained control from a single long caption is therefore fundamentally an attention routing problem---the goal is not richer conditioning, but selective access to the relevant part of the specification while retaining the acoustic history needed for autoregressive continuity.


Motivated by this observation, we propose MusicSTAR, which turns a single long caption into temporally selective conditioning through three complementary components. First, the Section-wise Conditioner factorizes caption and lyric representations at the section level and isolates interactions within the conditioning prefix, preventing cross-section information from being mixed before routing. Second, the Temporal Attention Scheduler learns to advance a monotonic section pointer from the generation state itself, enabling adaptive transitions without explicit timestamps or durations. Third, Temporal Attention Redistribution shifts caption attention toward the active section while exactly preserving the attention mass assigned to audio history, strengthening temporal specificity without disrupting autoregressive continuity. Together, these components isolate, schedule, and route temporal conditions within a unified autoregressive generation process.


Two training elements support this design. Dual-view training reduces over-reliance on automatically inferred section boundaries, and a local-to-global preference strategy, which contains Section-DPO for section-level controllability followed by Song-DPO for whole-song musicality, aligns preference learning with the temporal granularity of control. Experiments show that MusicSTAR substantially improves fine-grained temporal instruction following over the strongest section-level control baseline while matching the best open-weight systems in overall song quality.


Our main contributions are:
\begin{itemize}
    \item We introduce a structured, robust, and extensible temporal control representation that encodes a song's global characteristics and complete section-wise evolution in a single long caption.
    \item We propose a structure-aware temporal attention routing framework comprising a Section-wise Conditioner, a Temporal Attention Scheduler, and Temporal Attention Redistribution, which respectively isolate, schedule, and route section-level conditions within a unified autoregressive generation process.
    \item We introduce a local-to-global preference optimization strategy that aligns preference learning with the temporal granularity of control, progressing from section-level controllability with Section-DPO to whole-song musicality with Song-DPO.
\end{itemize}


\section{Related Work}
\subsection{Song Generation}
Recent song generation systems have advanced both autoregressive and diffusion-based approaches for lyrics and text-conditioned synthesis. Seed-Music~\cite{bai2024seed} introduces a unified framework that leverages autoregressive language modeling and diffusion-based approaches to support high-quality, fine-grained controllable song generation. YuE~\cite{yuan2026yue}, SongCreator~\cite{lei2024songcreator}, MiniMax-Music3~\cite{minimaxmusic2026} and Qwen-Music~\cite{xu2026qwen} explored autoregressive lyrics-based generation, while LeVo~\cite{lei2026levo}, HeartMuLa~\cite{yang2026heartmula} established a strong autoregressive framework for high-quality song generation with mixed and track-specific audio representations. ACE-Step 1.5~\cite{gong2026ace} and DiffRhythm 2~\cite{jiang2025diffrhythm} employ diffusion Transformer-based architectures to enable coherent and controllable music generation. These models substantially improve full-song synthesis, but primarily treat textual conditions as global or sequence-level context rather than as temporally routed conditions.

\subsection{Temporal Control}
Beyond global conditioning, fine-grained music control has increasingly incorporated time-varying signals, as in Music ControlNet~\cite{wu2024music} and Mustango~\cite{melechovsky2024mustango}. For song generation, SegTune~\cite{wang2026segtune} and Muse~\cite{jiang2026muse} are the closest approaches, externalizing temporal structure through predicted condition-to-time alignment and a segment-structured generation format, respectively. MusicSTAR differs in \emph{where} temporal structure lives: the entire evolution is expressed implicitly in a single long caption and resolved internally through attention routing during unified autoregressive generation.

\subsection{Preference Alignment}
Preference optimization provides another axis for improving song generation. LeVo introduces multi-preference DPO~\cite{rafailov2023direct}, and LeVo 2~\cite{lei2026levo} further adopts progressive post-training for generation quality, controllability, and musicality. These approaches primarily formulate preference optimization at the whole-song level. MusicSTAR instead matches preference learning to the temporal granularity of control, progressing from section-level controllability with Section-DPO to whole-song musicality with Song-DPO.

\begin{figure}[t]
\begin{center}
\includegraphics[width=\linewidth]{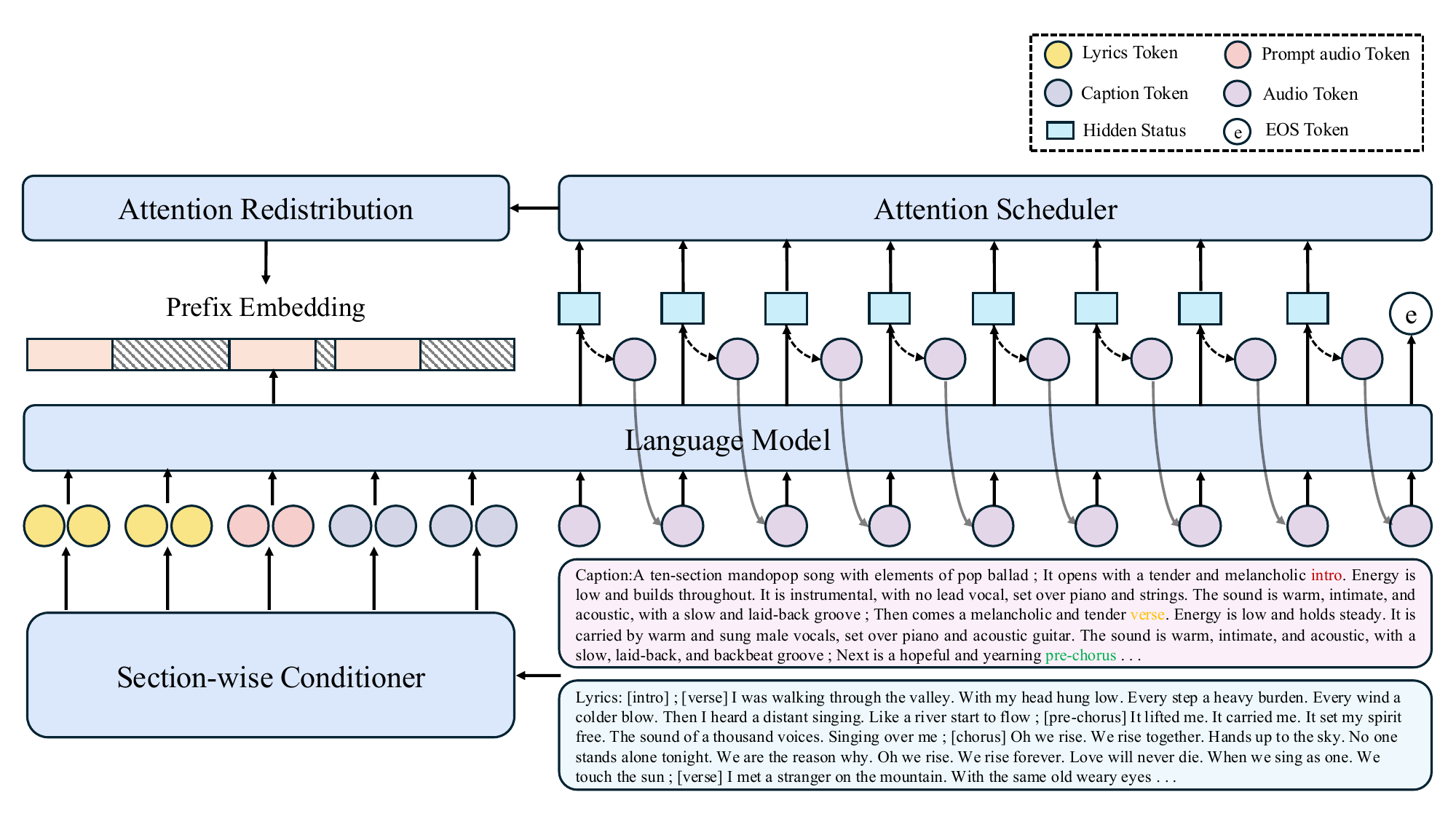}
\end{center}
\caption{Overview of MusicSTAR. The Section-wise Conditioner transforms the Single Long Caption and lyrics into a section-isolated prefix. During autoregressive generation, the Temporal Attention Scheduler updates the active section across steps, while Temporal Attention Redistribution focuses attention on the corresponding conditioning content without disrupting access to audio history. This cross-step routing enables timestamp-free temporal control from a single long caption.}
\end{figure}

\section{Method}
MusicSTAR follows a simple principle: separate temporal conditions, route them during generation, and align generation from local control to global quality. We first encode the complete temporal evolution in a structured Single Long Caption with section-pure conditioning, then apply Temporal Attention Routing to dynamically focus on the relevant section and schedule temporal transitions, followed by boundary-robust training and local-to-global preference optimization.

\subsection{Structured Single Long Caption}

We formulate song generation with a Single Long Caption that specifies the global characteristics and complete temporal evolution of the specified musical attributes in a single textual condition. Let a song contain $N$ ordered sections with acoustic spans
\begin{equation}
B^*=\{[s_i,e_i)\}_{i=1}^{N},
\label{eq:section_spans}
\end{equation}
and represent the caption as
\begin{equation}
C=[g;c_1;c_2;\ldots;c_N],
\label{eq:caption}
\end{equation}
where $g$ denotes the global description and $c_i$ describes the $i$-th section. For each section, a fixed set of user-facing musical attributes is normalized and fused into natural language. The resulting clauses are ordered by the song structure and preceded by the global description, forming a unified global-to-local specification without explicit timestamps or durations. Controlled attribute dropout further reduces reliance on rigid attribute-to-text correspondences while preserving the underlying structure, making the representation robust and extensible to different attribute schemas.

Given lyrics $L$ and caption $C$, the model generates acoustic tokens autoregressively:
\begin{equation}
x_t \sim P_{\theta}(x_t \mid x_{<t},L,C).
\label{eq:autoregressive}
\end{equation}
Although the complete temporal specification remains accessible throughout generation, the condition relevant to the current musical state should dominate at each step. Let $i_t$ denote the active section at step $t$. Ideally,
\begin{equation}
x_t \sim P_{\theta}(x_t \mid x_{<t},L,g,c_{i_t}),
\label{eq:target_conditioning}
\end{equation}
while retaining access to globally relevant conditions and previously generated audio. This reveals the central challenge of the single-caption setting: \emph{the temporal information is fully specified, but its access is not temporally selective}. MusicSTAR addresses this gap by first constructing section-pure conditioning representations and then applying Temporal Attention Routing to determine when the active section changes and which temporal condition should receive attention.

\subsection{Section-wise Conditioner}
Effective temporal routing requires section-specific information to remain distinguishable throughout conditioning. Jointly encoding the full caption and lyrics can introduce cross-section dependencies before routing. We therefore introduce the Section-wise Conditioner, which factorizes section representations and isolates information flow within the conditioning prefix.

Let $l_i$ denote the lyrics of section $i$. We independently encode the global description, each section caption, and each lyric section, and reassemble their representations into a unified prefix:
\begin{equation}
P=
\operatorname{Reassemble}
\left(
E_g(g),
\{E_c(c_i)\}_{i=1}^{N},
\{E_l(l_i)\}_{i=1}^{N}
\right),
\label{eq:sectionwise_conditioner}
\end{equation}
where $E_g$, $E_c$, and $E_l$ independently encode global, section-level caption, and section-level lyric content, respectively. Although $P$ is a single continuous prefix, each section representation is computed independently of other sections, providing section-pure conditioning for subsequent routing. To prevent cross-section mixing after reassembly, each prefix token $k$ is assigned a region index $R(k)\in\{-1,1,\ldots,N\}$, where $R(k)=-1$ denotes globally visible content and $R(k)=i$ denotes the caption and lyrics of section $i$. We impose the hard prefix mask
\begin{equation}
M^{\mathrm{pre}}_{qk}=
\begin{cases}
0, & R(q)=R(k)\ \text{or}\ R(k)=-1,\\
-\infty, & \text{otherwise},
\end{cases}
\label{eq:prefix_isolation}
\end{equation}
so that section-specific tokens interact only within their own section or with global content, while global tokens cannot serve as cross-section information relays. This isolation is always active and independent of the temporal routing strength.

\subsection{Temporal Attention Routing}
The Section-wise Conditioner separates temporal conditions but does not determine which condition should control generation at a particular step. MusicSTAR therefore introduces Temporal Attention Routing, which decomposes temporal control into two complementary functions: the Temporal Attention Scheduler determines \emph{when to transition}, while the Temporal Attention Redistribution determines \emph{where to attend}.

\subsubsection{Temporal Attention Scheduler}
The Temporal Attention Scheduler enables adaptive section progression without explicit timestamps or durations at inference time. Let $h_t$ denote the final hidden state at the current audio query position. A lightweight boundary head predicts whether the current section terminates at frame $t$:
\begin{equation}
p_t=\sigma(w^\top h_t+b).
\label{eq:boundary_prediction}
\end{equation}
The scheduler maintains a monotonic section pointer. If the predicted boundary exceeds a threshold $\tau$ and the current section has lasted at least $L_{\min}$ frames, the pointer advances:
\begin{equation}
i_{t+1}=
\begin{cases}
i_t+1, & p_t>\tau\ \land\ t-t_i\geq L_{\min},\\
i_t, & \text{otherwise},
\end{cases}
\label{eq:scheduler}
\end{equation}
where $t_i$ is the start frame of the current section.

For a section span $[s_i,e_i)$, we assign the boundary target to its final frame, such that a boundary detected at frame $t$ activates the next section from frame $t+1$ onward. This convention keeps the temporal indexing consistent between boundary supervision and inference-time section transitions.

\subsubsection{Temporal Attention Redistribution}
Once the active section is determined, the model should preferentially access its corresponding caption and lyric information while retaining global conditions, prompt audio, and generated audio history. A naive suppression strategy reduces attention to irrelevant sections, but also changes the overall conditioning--history attention balance. We instead redistribute attention within the prefix.

Let $s_{t,k}$ denote the attention logit from audio query $t$ to prefix key $k$. For the current routing state, let $\mathcal{P}_t$ denote the preferred prefix region, consisting of globally visible conditions, prompt audio, and the caption and lyrics associated with the active section. The remaining section-specific prefix tokens receive a soft penalty:
\begin{equation}
\delta_{t,k}=
\begin{cases}
0, & k\in\mathcal{P}_t,\\
-\lambda_t, & k\in\mathcal{P}\setminus\mathcal{P}_t,
\end{cases}
\label{eq:routing_bias}
\end{equation}
where $\mathcal{P}$ denotes the set of prefix keys visible to the query.

In the mass-conserving formulation, we compensate this reweighting with a row-adaptive normalization term and obtain
\begin{equation}
\widetilde{s}_{t,k}
=
s_{t,k}
+\delta_{t,k}
+
\log
\frac{
\sum_{j\in\mathcal{P}}\exp(s_{t,j})
}{
\sum_{j\in\mathcal{P}}\exp(s_{t,j}+\delta_{t,j})
},
\qquad k\in\mathcal{P}.
\label{eq:redistribution}
\end{equation}
This construction preserves the total prefix attention mass:
\begin{equation}
\sum_{k\in\mathcal{P}}\exp(\widetilde{s}_{t,k})
=
\sum_{k\in\mathcal{P}}\exp(s_{t,k}).
\label{eq:mass_conservation}
\end{equation}
Because audio-history logits remain unchanged, the attention allocated to the generated history is preserved as well. Temporal control is therefore achieved by changing the relative allocation within the conditioning prefix, rather than by trading conditioning attention against autoregressive history.

The scheduler and redistribution consequently play distinct and complementary roles. The former determines the temporal state of generation, while the latter realizes that state through selective attention over the structured prefix. When the scheduler advances to the next section, the preferred prefix region changes on the following generation step, causing attention to progress through the ordered section descriptions.

\subsection{Boundary-Robust Training}
Section boundaries obtained from automatic music-structure analysis are inherently uncertain, making routing sensitive to small boundary shifts. We therefore introduce a dual-view training strategy that alternates between the exact boundary set $B^*$ and a perturbed view $\widetilde{B}$. Specifically, each internal boundary is independently shifted by a Laplace-distributed offset, which is shared by the two adjacent sections to preserve temporal continuity. The perturbation scale is set to 25 frames in our default training configuration.

For each training sample, one view is selected stochastically:
\begin{equation}
B=
\begin{cases}
B^*, & \text{with probability } p_{\mathrm{exact}},\\
\widetilde{B}, & \text{otherwise}.
\end{cases}
\label{eq:dual_view}
\end{equation}
This dual-view training exposes the model to realistic variations in section boundaries, improving the robustness of temporal routing to boundary localization errors while maintaining reliable temporal supervision.

\subsection{Local-to-Global Preference Optimization}
Supervised training establishes the correspondence between the structured caption and temporal generation, but preference optimization requires feedback at different temporal granularities. A whole-song preference signal can obscure local control failures, as a poorly controlled section may be averaged with otherwise satisfactory portions of the song. We therefore optimize preferences progressively from individual sections to the entire song, aligning the supervision region with the level at which quality is evaluated.

For a preference region $R$, we use a length-normalized log-likelihood
\begin{equation}
\bar{\ell}_{\theta}(y\mid R)
=
\frac{1}{|R|}
\sum_{t\in R}
\log \pi_{\theta}(y_t\mid y_{<t},L,C),
\label{eq:normalized_loglikelihood}
\end{equation}
and apply the standard DPO objective
\begin{equation}
\mathcal{L}_{\mathrm{DPO}}
=
-\log \sigma
\left(
\beta
\left[
\bar{\ell}_{\theta}(y^+\mid R)
-
\bar{\ell}_{\theta}(y^-\mid R)
-
\bar{\ell}_{\mathrm{ref}}(y^+\mid R)
+
\bar{\ell}_{\mathrm{ref}}(y^-\mid R)
\right]
\right).
\label{eq:dpo}
\end{equation}

In the first stage, Section-DPO restricts $R$ to the acoustic span of a target section, with preferences determined by local controllability. Candidates are compared under the same song and section context, using Mulan similarity to measure adherence to the prescribed musical attributes. Because the preference objective is evaluated only on the target section, the learning signal is localized to the region where control is deficient, rather than being diluted by already satisfactory sections elsewhere in the song. This enables targeted refinement of individual sections while preserving the full-song context used for generation.

After Section-DPO, we resample candidate songs from the resulting checkpoint and perform Song-DPO with $R$ covering the entire song. Preference is then based on whole-song musicality measured by SongBench, allowing the model to refine qualities that emerge only at the global level, such as overall coherence and musical quality.

The two stages share the same DPO objective but differ in supervision region and preference signal. They are applied strictly sequentially, with Song-DPO initialized from the Section-DPO model and trained on newly generated candidates. This local-to-global strategy therefore follows the structure of the control problem itself: first refine specific temporal conditions where they fail, then optimize the quality and coherence of the complete song.

\begin{table}[t]
\caption{Quantitative comparison on AudioBox Aesthetics and SongBench. Bold values denote the best results, while underlined values denote the second-best results.}
\label{tab:quantitative_results}
\begin{center}
\scriptsize
\setlength{\tabcolsep}{2.0pt}
\renewcommand{\arraystretch}{1.05}
\resizebox{\columnwidth}{!}{
\begin{tabular}{lccccc|cccccccc}
\hline
\multicolumn{1}{c}{\bf Method}
& \multicolumn{5}{c|}{\bf AudioBox Aesthetics}
& \multicolumn{8}{c}{\bf SongBench} \\
\cline{2-14}
& \bf CE & \bf CU & \bf PC & \bf PQ & \bf Mean
& \bf Mel. & \bf Arr. & \bf Mus. & \bf Voc. & \bf Inst. & \bf Mix. & \bf Str. & \bf Mean \\
\hline
Muse
& 7.55 & 7.75 & \bf 6.60 & 8.21 & 7.53
& 6.19 & 6.23 & 5.26 & 6.45 & 6.40 & 6.21 & 6.02 & 6.11 \\

ACE-Step 1.5
& \underline{7.68} & 7.91 & 6.18 & 8.26 & 7.51
& 6.33 & 6.56 & 5.27 & 6.30 & 6.64 & 6.32 & 6.26 & 6.24 \\

HeartMuLa
& 7.66 & 7.91 & 6.46 & 8.27 & 7.57
& 6.41 & 6.47 & 5.61 & \underline{6.88} & 6.58 & 6.50 & 6.32 & 6.40 \\

LeVo 2
& \underline{7.68} & \bf 7.98 & \underline{6.53} & \bf 8.41 & \underline{7.65}
& 6.50 & \underline{6.84} & \underline{5.66} & 6.79 & \bf 6.93 & \underline{6.59} & 6.32 & 6.52 \\

MiniMax-Music3
& 7.54 & 7.86 & 6.43 & 8.19 & 7.51
& \bf 6.58 & 6.82 & 5.61 & \bf 6.89 & 6.78 & \bf 6.61 & \bf 6.60 & \bf 6.56 \\

MusicSTAR
& \bf 7.75 & \underline{7.96} & 6.52 & \underline{8.39} & \bf 7.66
& \underline{6.54} & \bf 6.88 & \bf 5.72 & 6.76 & \underline{6.89} & \underline{6.59} & \underline{6.40} & \underline{6.54} \\
\hline
\end{tabular}
}
\end{center}
\end{table}

\begin{table}[t]
\caption{Fine-grained control analysis between MusicSTAR and Muse using MuLan similarity under two protocols. Column abbreviations: Eng.-L (Energy Level), Eng.-A (Energy Arc), Voc.-L (Vocal Lead), Voc.-E (Vocal Ensemble), Inst. (Instrumentation), Style (Sound Style). Bold values denote the better result in each column.}
\label{tab:mulan_analysis}
\begin{center}
\scriptsize
\setlength{\tabcolsep}{1.8pt}
\renewcommand{\arraystretch}{1.03}
\resizebox{\columnwidth}{!}{
\begin{tabular}{llcccccccc}
\hline
\multicolumn{1}{c}{\bf Eval.}
& \multicolumn{1}{c}{\bf Method}
& \bf Eng.-L & \bf Eng.-A & \bf Mood & \bf Voc.-L
& \bf Voc.-E & \bf Inst. & \bf Style & \bf Rhythm \\
\hline
Direct
& Muse
& \bf 0.1342 & 0.1718 & 0.1490 & 0.3640 & 0.2262 & 0.1137 & \bf 0.1378 & 0.1411 \\

Direct
& MusicSTAR
& 0.1332 & \bf 0.1879 & \bf 0.1737 & \bf 0.3681
& \bf 0.2283 & \bf 0.1721 & 0.1372 & \bf 0.1457 \\
\hline
Extracted
& Muse
& 0.7977 & 0.8677 & 0.6128 & 0.6135 & 0.8554 & 0.2920 & 0.5877 & 0.7415 \\

Extracted
& MusicSTAR
& \bf 0.8207 & \bf 0.8721 & \bf 0.6357 & \bf 0.6800
& \bf 0.8578 & \bf 0.4387 & \bf 0.6234 & \bf 0.7578 \\
\hline
\end{tabular}
}
\end{center}
\end{table}

\section{Experiments}    
\subsection{Baselines.}
We compare MusicSTAR against both the most directly related approach for fine-grained song control and strong open-weight music generation models. Muse is the closest baseline to our setting, as it supports section-level style control within long-form song generation and provides a fully reproducible open-source implementation. We do not include SegTune, another closely related temporal-control approach, because its pretrained weights are not publicly available for direct reproduction in our evaluation setting.

To assess MusicSTAR against strong general-purpose song generation systems, we further include LeVo 2, MiniMax-Music 3, ACE-Step 1.5, and HeartMuLa. These models cover recent high-performing open-weight systems with different generation paradigms and conditioning mechanisms, providing a broad comparison beyond methods specifically designed for temporal control. For all baselines, we use their publicly released checkpoints and official inference procedures whenever available, and follow their recommended settings as closely as possible.

\subsection{Evaluation Metrics}
We evaluate generated songs from both objective and subjective perspectives. For objective evaluation, we employ AudioBox Aesthetics~\cite{tjandra2025meta} and SongBench~\cite{wu2026songbench}, covering complementary aspects of perceptual quality and musical quality. AudioBox Aesthetics evaluates general audio quality across content and production dimensions, while SongBench provides fine-grained assessment of song-specific musical attributes.

For subjective evaluation, we assess six dimensions that directly reflect generation quality and the proposed control objective: Overall Musical Quality, Vocal \& Accompaniment Quality, Vocal-Accompaniment Harmony, Song Structure Clarity, Lyric Fidelity, and Instruction Adherence. This evaluation jointly captures general song quality and fine-grained controllability.

\subsection{Quantitative Results}
We first evaluate overall song quality in Table~\ref{tab:quantitative_results}. MusicSTAR attains the best AudioBox mean, indicating that structured temporal conditioning and attention routing do not compromise general audio quality, and reaches a SongBench mean of 6.54, substantially above Muse and on par with the strongest open-weight systems. Overall song quality and fine-grained temporal control are therefore jointly achievable within a single unified generation process.

Song-level metrics, however, do not reveal whether the specified attributes are realized in the intended sections. We therefore conduct a dedicated eight-dimensional control analysis against Muse, the only baseline supporting section-level control explicitly, under two MuLan protocols (Table~\ref{tab:mulan_analysis}): \emph{Direct} computes audio-to-text similarity between generated section audio and the ground-truth attribute descriptions, whereas \emph{Extracted} first extracts section-level attributes from the generated audio and then measures text-to-text similarity. For fairness, both systems are driven by the same section-level attribute descriptions, converted into each model's native conditioning format.

Under the Extracted protocol, MusicSTAR leads on all eight dimensions, with the largest margins on Vocal Lead and Instrumentation. Under the Direct protocol, differences are smaller and less consistent, confirming that raw audio-text similarity is relatively insensitive to fine-grained control attributes. These attribute-level gains are further corroborated by the human instruction-adherence evaluation reported below. We thus arrive at the central empirical message of this work: fine-grained temporal control can be improved substantially without sacrificing overall song quality.

\subsection{Subjective Results}

\begin{table}[t]
\caption{Subjective evaluation by mean opinion score on a five-point scale, covering the six dimensions defined in the evaluation metrics. Bold values denote the best results, while underlined values denote the second-best results.}
\label{tab:subjective}
\begin{center}
\small
\setlength{\tabcolsep}{4.0pt}
\renewcommand{\arraystretch}{1.05}
\begin{tabular}{lccccccc}
\hline
\multicolumn{1}{c}{\bf Method}
& \bf Quality & \bf Voc.\&Acc. & \bf Harmony & \bf Structure
& \bf Lyrics & \bf Adherence & \bf Mean \\
\hline
Muse
& 3.23 & 3.34 & 3.44 & 3.39 & 3.53 & 3.11 & 3.34 \\

ACE-Step 1.5
& 3.61 & 3.66 & \underline{3.79} & 3.66 & 3.69 & \underline{3.59} & 3.66 \\

HeartMuLa
& 3.53 & 3.54 & 3.51 & 3.61 & 3.77 & 3.46 & 3.57 \\

LeVo 2
& \underline{3.80} & \underline{3.67} & 3.76 & \underline{3.67} & \underline{3.90} & 3.54 & \underline{3.72} \\

MiniMax-Music3
& 3.51 & 3.61 & 3.64 & 3.60 & 3.80 & 3.44 & 3.60 \\

MusicSTAR
& \bf 4.04 & \bf 4.01 & \bf 3.96 & \bf 4.00 & \bf 4.04 & \bf 3.99 & \bf 4.01 \\
\hline
\end{tabular}
\end{center}
\end{table}

We further conduct a human evaluation, in which annotators rate each generated song on the six dimensions using a five-point scale over the same held-out songs. As shown in Table~\ref{tab:subjective}, MusicSTAR attains the highest MOS on every dimension, with a clear margin in the overall mean. The gains are smallest on Harmony and Lyric Fidelity and largest precisely where temporal control operates, Instruction Adherence and Song Structure Clarity, mirroring the section-level controllability that the router is designed to provide. The advantage is more pronounced over the control-oriented Muse, showing that the objective improvements in Table~\ref{tab:mulan_analysis} are genuinely perceptible to listeners.

\subsection{Ablation Studies}
We ablate the two central claims of MusicSTAR---temporal control requires conditions to be separated, routed, and scheduled, and preference optimization should match the granularity of control. Component ablations use the SFT model, preference ablations the full architecture. \emph{Ctrl.} denotes the eight-dimensional Extracted MuLan similarity of Table~\ref{tab:mulan_analysis}.

\begin{table}[t]
\caption{Component ablation on the SFT model. \emph{w/o Scheduler} drives transitions with explicitly provided timestamps at inference. \emph{w/o Redistribution} removes the routing mechanism together with the conditioner. \emph{w/o Section-wise Conditioner} jointly encodes caption and lyrics without factorization or isolation, with scheduling and redistribution applied to the entangled section spans.}
\label{tab:component_ablation}
\begin{center}
\scriptsize
\setlength{\tabcolsep}{2.0pt}
\renewcommand{\arraystretch}{1.05}
\resizebox{\columnwidth}{!}{
\begin{tabular}{lccccc|cccccccc}
\hline
\multicolumn{1}{c}{\bf Variant}
& \multicolumn{5}{c|}{\bf AudioBox Aesthetics}
& \multicolumn{8}{c}{\bf SongBench} \\
\cline{2-14}
& \bf CE & \bf CU & \bf PC & \bf PQ & \bf Mean
& \bf Mel. & \bf Arr. & \bf Mus. & \bf Voc. & \bf Inst. & \bf Mix. & \bf Str. & \bf Mean \\
\hline
MusicSTAR (SFT)
& \bf 7.76 & \bf 7.97 & \underline{6.41} & \bf 8.41 & \bf 7.64
& \bf 6.15 & \bf 6.42 & \bf 5.35 & \bf 6.42 & \bf 6.61 & \bf 6.27 & \bf 6.08 & \bf 6.19 \\

\quad w/o Scheduler
& 7.74 & \bf 7.97 & \bf 6.43 & \bf 8.41 & \bf 7.64
& \underline{6.03} & \underline{6.32} & \underline{5.19} & \underline{6.28} & \underline{6.51} & \underline{6.16} & \underline{5.98} & \underline{6.07} \\

\quad w/o Redistribution
& \underline{7.75} & \bf 7.97 & 6.29 & \underline{8.38} & \underline{7.60}
& 5.93 & 6.20 & 5.10 & 6.25 & 6.45 & 6.07 & 5.91 & 5.99 \\

\quad w/o Section-wise Conditioner
& 7.41 & \underline{7.90} & 6.07 & 8.20 & 7.39
& 5.47 & 5.89 & 4.52 & 5.45 & 6.09 & 5.61 & 5.37 & 5.49 \\
\hline
\end{tabular}
}
\end{center}
\end{table}

\begin{table}[t]
\caption{Ablation of preference optimization schedules on the full architecture. \emph{Section-only} and \emph{Song-only} apply a single stage on the SFT model; \emph{Mixed} trains one stage on the union of both pair types under the same budget; \emph{Section\,$\to$\,Song} is the proposed schedule.}
\label{tab:dpo_ablation}
\begin{center}
\scriptsize
\setlength{\tabcolsep}{1.9pt}
\renewcommand{\arraystretch}{1.05}
\resizebox{\columnwidth}{!}{
\begin{tabular}{lcccccc|cccccccc}
\hline
\multicolumn{1}{c}{\bf Variant}
& \multicolumn{1}{c}{\bf Ctrl.}
& \multicolumn{5}{c|}{\bf AudioBox Aesthetics}
& \multicolumn{8}{c}{\bf SongBench} \\
\cline{2-15}
& \bf 8-dim
& \bf CE & \bf CU & \bf PC & \bf PQ & \bf Mean
& \bf Mel. & \bf Arr. & \bf Mus. & \bf Voc. & \bf Inst. & \bf Mix. & \bf Str. & \bf Mean \\
\hline
SFT
& 0.7002
& 7.76 & 7.97 & 6.41 & 8.41 & 7.64
& 6.15 & 6.42 & 5.35 & 6.42 & 6.61 & 6.27 & 6.08 & 6.19 \\

Section-only
& 0.7025
& \underline{7.78} & \underline{7.98} & 6.44 & \underline{8.42} & \underline{7.66}
& 6.14 & 6.43 & 5.32 & 6.41 & 6.62 & 6.27 & 6.07 & 6.18 \\

Song-only
& 0.7055
& \bf 7.81 & \bf 7.99 & \bf 6.52 & \bf 8.43 & \bf 7.69
& \underline{6.33} & \underline{6.65} & \underline{5.53} & \underline{6.56} & \underline{6.75} & \underline{6.47} & \underline{6.23} & \underline{6.36} \\

Mixed
& \underline{0.7084}
& 7.76 & \underline{7.98} & \underline{6.48} & \underline{8.42} & \underline{7.66}
& 6.14 & 6.43 & 5.31 & 6.46 & 6.61 & 6.25 & 6.08 & 6.18 \\

Section\,$\to$\,Song 
& \bf 0.7105
& 7.75 & 7.96 & \bf 6.52 & 8.39 & \underline{7.66}
& \bf 6.54 & \bf 6.88 & \bf 5.72 & \bf 6.76 & \bf 6.89 & \bf 6.59 & \bf 6.40 & \bf 6.54 \\
\hline
\end{tabular}
}
\end{center}
\end{table}

Table~\ref{tab:component_ablation} isolates each design. Removing the routing mechanism while keeping the same caption (\emph{w/o Redistribution}) degrades musical quality but not aesthetics, directly quantifying temporal condition misrouting: the deficit lies in \emph{how} the condition is accessed, not \emph{what} is specified. Explicit timestamps at inference (\emph{w/o Scheduler}) do not outperform the learned scheduler, which shows timestamp-free control is not a compromise, since the scheduler advances with the generated content, whereas the provided timestamps inherit the boundary noise of automatic structure analysis. Routing over jointly encoded conditions (\emph{w/o Section-wise Conditioner}) is worse than no routing at all: once sections are mixed, the routing bias amplifies contamination rather than restoring specificity, with the damage concentrating on musical organization rather than surface quality. 

Table~\ref{tab:dpo_ablation} varies the granularity and ordering of preference supervision. No single granularity suffices on its own, and mixing both in one stage only makes them interfere; sequencing converts interference into synergy: \emph{Section\,$\to$\,Song} attains the best musicality across all SongBench dimensions and the best controllability at a negligible aesthetic cost, as Section-DPO first stabilizes local control so that Song-DPO can optimize global coherence rather than compensate for local failures.

\section{Conclusion}
We presented MusicSTAR, which turns a single long caption into temporally selective conditioning for song generation: by isolating, scheduling, and routing section-level conditions, it resolves temporal condition misrouting, the gap between a fully specified caption and selective access to it. MusicSTAR substantially improves fine-grained temporal instruction following while matching the strongest open-weight systems in overall song quality, and its learned scheduler even rivals explicit timestamp alignment. Our results suggest that effective temporal control depends not only on what is specified, but also on how it is accessed.

\subsection*{AI USE STATEMENT}
In this work, we used generative AI tools for (i) aiding and polishing writing, (ii) drafting selected sections of the paper, (iii) retrieval and discovery of related work, and (iv) automatic annotation of large-scale pretraining data. 

We have not used generative AI tools to formulate research ideas or hypotheses, to develop theoretical or conceptual frameworks or mathematical claims, to design the methodology or experiments, to implement the proposed method, or to process, analyze, or interpret experimental results. All other tasks with required disclosure are not applicable to this work.

We have reviewed all AI-assisted work. All AI-drafted and AI-polished text was manually reviewed, verified, and edited by the authors. The AI-generated pretraining annotations were validated by human inspection on randomly sampled subsets, together with automatic consistency checks against the source captions. All technical descriptions, experimental results, numerical values, and claims were independently verified by the authors against the implementation and experimental records; no experimental results, numerical values, or references were produced by generative AI without independent verification. We take responsibility for the final content of this work, including text, claims, and artifacts produced with the aid of generative AI.

\subsection*{Ethics statement}
The subjective evaluation was conducted with informed consent from all annotators. The training corpus is internally licensed for research use. As with any generative music system, MusicSTAR may be misused to produce misleading or infringing content; we encourage deployment with appropriate content attribution mechanisms.

\subsection*{Reproducibility statement}
We describe the architecture and training objectives in Section~3, the dataset construction, preference-pair collection and training configurations in Appendix. All baselines are evaluated using publicly released checkpoints and official inference settings, and all evaluation metrics are computed with publicly available models. Audio samples are available at \url{https://anonymous.4open.science/w/ICLR_30705_demopage/}, and the source code will be released upon publication.

\bibliography{iclr2027_conference}
\bibliographystyle{iclr2027_conference}

\appendix
\section{Dataset}
We conduct experiments on an internal dataset containing approximately 1 million songs. To construct the proposed Single Long Caption, we first apply SongFormer~\cite{hao2025songformer} to segment each song into ordered musical sections, providing temporal structure for subsequent caption construction. For each section, Qwen3-Omni-30B-A3B-Instruct~\cite{xu2025qwen3} is used to extract eight user-perceivable and controllable musical attributes. The section-level attributes are then fused into natural-language descriptions according to the inferred song structure, with a dropout rate of $0.1$ during fusion. This pipeline combines explicit structural organization with linguistic robustness, while remaining extensible: the underlying attribute schema can be flexibly expanded or reduced for different application domains without changing the overall captioning framework.

For preference optimization, we construct training pairs in two stages corresponding to the local-to-global optimization strategy. In the first stage, we generate candidate songs, extract section-level attributes from their audio using Qwen3-Omni-30B-A3B-Instruct, and compare the predicted attributes with the ground-truth attributes using MuQ-MuLan~\cite{zhu2025muq} text-to-text similarity. We use text-to-text comparison rather than directly relying on audio-text similarity, as the latter is less sensitive to fine-grained attribute-level differences. This stage yields approximately 44K section-level preference pairs for Section-DPO. In the second stage, candidate songs are evaluated with SongBench, and approximately 16K whole-song preference pairs are constructed for Song-DPO. The two datasets therefore provide preference supervision at complementary temporal granularities, from section-level controllability to whole-song musicality.

\section{Implementation Details.}

We build MusicSTAR upon a from-scratch reproduction of the LeVo architecture, trained on our 1M-song dataset. The Sectionwise Conditioner is initialized from the publicly released Qwen3-Embedding-0.6B~\cite{zhang2025qwen3} and adopts the corresponding pretrained weights released with ACE-Step 1.5.

For supervised fine-tuning, we train the model for 30K steps with a batch size of 128 and a learning rate of $1\times10^{-4}$ on 64 NVIDIA H20 GPUs. We then perform Section-DPO for 2K steps with a batch size of 32 and a learning rate of $1\times10^{-6}$ on 32 NVIDIA H20 GPUs, followed by Song-DPO for 1K steps with a batch size of 16 and the same learning rate of $1\times10^{-6}$ on 16 NVIDIA H20 GPUs. The two preference optimization stages are conducted sequentially: Song-DPO is initialized from the Section-DPO checkpoint and further optimizes whole-song preferences after local section-level alignment. This training schedule progressively shifts the optimization target from supervised temporal conditioning, to section-level controllability, and finally to whole-song musicality.


\section{Eight-Dimensional Temporal Attribute Schema}
To construct the Single Long Caption, we define an eight-dimensional temporal attribute schema designed around user-perceivable and controllable musical characteristics. The schema focuses on properties that listeners can recognize and that can be naturally expressed as textual control instructions, rather than low-level or explicitly quantified musical descriptors. Each section is represented by a compact set of tags under fixed semantic and ordering constraints, which are subsequently fused into natural-language descriptions.
\begin{table}[t]
\caption{The eight-dimensional temporal attribute schema used to construct section-level descriptions.}
\label{tab:attribute_schema}
\begin{center}
\scriptsize
\setlength{\tabcolsep}{3.0pt}
\renewcommand{\arraystretch}{1.08}
\begin{tabular}{p{0.20\columnwidth}p{0.32\columnwidth}p{0.35\columnwidth}}
\hline
\textbf{Attribute} & \textbf{Function} & \textbf{Typical values / organization} \\
\hline
Energy Level
& Absolute intensity of a section, perceived through loudness, rhythmic drive, layering, and timbral weight
& very low / low / medium / high / very high \\

Energy Arc
& Within-section evolution of intensity, comparing the beginning, middle, and end
& steady / building / fading / peak in middle / dip in middle / up and down \\

Mood
& Perceived emotional character based on timbre, harmonic color, and vocal expression
& e.g., melancholic, euphoric, tense, dreamy, tender, ominous, playful \\

Vocal Lead
& Character and delivery of the lead vocal
& character $\rightarrow$ delivery $\rightarrow$ voice type; e.g., breathy, softly sung, male \\

Vocal Ensemble
& All vocal components beyond the lead vocal
& harmony, group vocals, doubling, independent vocal parts, interaction, and ornamentation \\

Instrumentation
& Audible instrumental content and its dominant timbral sources
& e.g., distorted electric guitar, 808 kick, electric piano, smooth bass \\

Sound Style
& Overall sonic texture, aesthetic character, and source type
& texture $\rightarrow$ aesthetic $\rightarrow$ source; e.g., warm, intimate, electronic \\

Rhythm Feel
& Perceived pace and rhythmic groove
& pace $\rightarrow$ groove; e.g., slow, laid-back, behind the beat \\
\hline
\end{tabular}
\end{center}
\end{table}

The schema is deliberately defined at the level of \emph{perception} rather than low-level measurement. In particular, it does not require BPM, key, time signature, chord names, note-level information, frequency measurements, or other quantitative descriptors. Rhythm is expressed through perceptual pace and groove terms, while sound style captures audible texture and aesthetic character without inferring properties such as lo-fi quality from missing high-frequency content. Likewise, the attribute values are inferred solely from the input audio.

Each dimension follows a constrained vocabulary or ordering rule where appropriate. Energy Level uses a closed five-level scale, whereas Energy Arc captures relative within-section change. Mood is represented by a small set of salient descriptors ordered by importance. Vocal Lead follows the order of character, delivery, and voice type, while Vocal Ensemble organizes non-lead vocal elements by functional category. Instrumentation prioritizes the most prominent audible instruments and excludes vocals. Sound Style follows the order of texture, aesthetic, and source, and Rhythm Feel places perceived pace before groove. These ordering conventions allow the tags to be directly realized as natural-language noun or adjective phrases while preserving their functional semantics.

We distinguish two forms of missing information. The value \texttt{none} indicates that an attribute is genuinely absent, such as no vocal content in a purely instrumental section, whereas an empty value indicates that the attribute cannot be reliably determined from the audio. This distinction avoids introducing arbitrary fallback labels and favors leaving uncertain attributes unspecified rather than guessing. We further apply automatic schema validation to enforce closed-set membership, tag-budget constraints, ordering rules, and category consistency during caption construction.

The schema is designed to be extensible. The eight dimensions constitute the current control vocabulary rather than a fixed ontology: new attributes can be added when a domain requires additional user-perceivable controls, while unnecessary dimensions can be removed without changing the overall caption construction procedure. The same global-to-local organization, section ordering, tag normalization, and natural-language fusion process therefore remains applicable across different control domains.

\section{Caption Routing and Global Coherence}
A natural concern is that discouraging attention to captions of inactive sections could sever cross-section dependencies, e.g., inconsistent melodies between repeated choruses or unresolved momentum from verse to chorus. We argue that such coherence is not primarily mediated by other sections' textual conditions, but by the acoustic history itself.

\begin{figure}[t]
\begin{center}
\includegraphics[width=\linewidth]{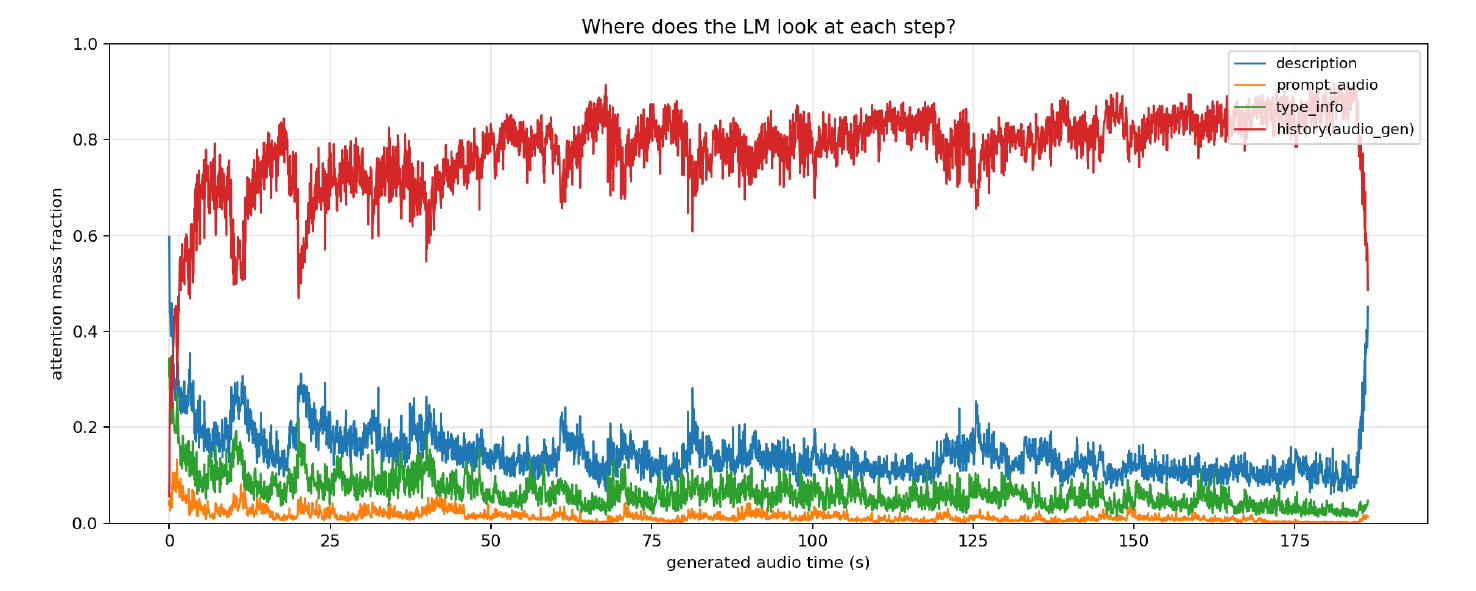}
\end{center}
\caption{Inference-time attention allocation of the baseline over one generated song, grouped by prefix role: descriptions (\texttt{lyrics}), prompt audio, type information (\texttt{caption}), and generated audio history (\texttt{history}). Audio history absorbs approximately 80\% of the attention mass throughout generation, whereas the caption occupies only a small residual share.}
\label{fig:attention_breakdown}
\end{figure}

Fig.~\ref{fig:attention_breakdown} visualizes the inference-time attention allocation of the baseline, with attention mass grouped by prefix role. Audio history dominates: it consistently absorbs roughly 80\% of the total attention mass across the whole song, while the caption receives only a small share. Melodic identity, timbre, groove, and transition dynamics are therefore read overwhelmingly from the model's own acoustic past. When a later chorus is generated, its melody remains consistent with the earlier one because the first chorus is literally present in the history; verse--chorus hand-offs are smoothed by the immediately preceding context window.

MusicSTAR's routing is designed to leave this channel intact by construction, in three complementary ways. First, Temporal Attention Redistribution operates only within the caption: the row-adaptive normalization in Eq.~\eqref{eq:redistribution} and Eq.~\eqref{eq:mass_conservation} conserves the total prefix attention mass, and audio-history logits are never modified, so the attention assigned to history is exactly what the model would allocate without routing. Second, the preferred prefix region $\mathcal{P}_t$ always contains the global description and prompt audio regardless of the routing state, so song-level style and timbre conditioning remain uniformly accessible at every step. Third, routing is soft rather than hard: inactive sections are attenuated by the finite penalty $-\lambda_t$ instead of being masked out, leaving residual access as a fallback whenever the acoustic evidence calls for it. In other words, control is steered by re-weighting \emph{which caption} guides the current step, while coordination and continuity continue to flow through \emph{where the model always listens}: its generated history and shared global conditions.

Empirically, the dimensions most sensitive to such disruption behave oppositely to the concern. MusicSTAR attains the top SongBench scores on Arrangement and Structure (Table~\ref{tab:quantitative_results}) and the highest subjective ratings on Vocal--Accompaniment Harmony and Song Structure Clarity (Table~\ref{tab:subjective}),precisely the metrics that would degrade first under cross-section incoherence.
\end{document}